\documentclass[letter,aps,showpacs,superscriptaddress,nofootinbib,tightenlines,prd]{revtex4}

\usepackage{amsfonts}
\usepackage{multirow}
\usepackage{mathrsfs}
\usepackage{graphicx}
\usepackage{amsmath}
\usepackage{amssymb}
\usepackage{bm}
\usepackage{bbm}
\usepackage{color}
\usepackage{ulem}
\usepackage{subfig}
\usepackage{caption}
\usepackage{graphicx}
\usepackage{epsfig}
\usepackage{braket}
\usepackage{slashed}
\usepackage{hyperref}
\usepackage{ulem}

\newcommand{\addnumber}{\addtocounter{equation}{1}\tag{\theequation}}
\newcommand{\msn}{{\pi}}
\newcommand{\bsq}{\begin{subequations}}
\newcommand{\esq}{\end{subequations}}

\def\Xint#1{\mathchoice
{\XXint\displaystyle\textstyle{#1}}%
{\XXint\textstyle\scriptstyle{#1}}%
{\XXint\scriptstyle\scriptscriptstyle{#1}}%
{\XXint\scriptscriptstyle\scriptscriptstyle{#1}}%
\!\int}
\def\XXint#1#2#3{{\setbox0=\hbox{$#1{#2#3}{\int}$ }
\vcenter{\hbox{$#2#3$ }}\kern-.6\wd0}}

\def\dashint{\Xint-}

\begin{document}

\title{\mbox{}\\[14pt]
Parton distributions of intrinsic charm in two-dimensional QCD}

\author{Siwei Hu~\footnote{husw@ihep.ac.cn}}
\affiliation{Institute of High Energy Physics, Chinese Academy of
	Sciences, Beijing 100049, China\vspace{0.2 cm}}
\affiliation{School of Physics, University of Chinese Academy of Sciences,
	Beijing 100049, China\vspace{0.2 cm}}

\author{Yu Jia~\footnote{jiay@ihep.ac.cn}}
\affiliation{Institute of High Energy Physics, Chinese Academy of
Sciences, Beijing 100049, China\vspace{0.2 cm}}
\affiliation{School of Physics, University of Chinese Academy of Sciences,
Beijing 100049, China\vspace{0.2 cm}}

\author{Zhewen Mo~\footnote{mozw@ihep.ac.cn}}
\affiliation{Institute of High Energy Physics, Chinese Academy of
Sciences, Beijing 100049, China\vspace{0.2 cm}}
\affiliation{School of Physics, University of Chinese Academy of Sciences,
Beijing 100049, China\vspace{0.2 cm}}

\author{Xiaonu Xiong~\footnote{xnxiong@csu.edu.cn}}
\affiliation{School of Physics and Electronics, Central South University, Changsha 418003, China}

\author{Mingliang Zhu~\footnote{lightzhu@csu.edu.cn}}
\affiliation{School of Physics and Electronics, Central South University, Changsha 418003, China}

\date{\today}

\begin{abstract}
We present a detailed investigation on the intrinsic charm content in a light meson  within the 't Hooft model,
namely, the two-dimensional QCD in large $N_c$ limit. The intrinsic charm  parton distribution function (PDF) of a light meson,
which first arises at order $N_c^{-1}$, is explicitly expressed in terms of the 't Hooft wave functions of the light meson
and an infinite tower of excited charmed mesons. We also derive the functional forms from the two-dimensional counterparts of
the meson cloud model (MCM) and Brodsky-Hoyer-Peterson-Sakai (BHPS) model. We then make a quantitative
comparison between our rigorous results and model predictions. We also study how the profile of the intrinsic charm PDF
varies with charm quark mass. The average momentum fraction carried by the charm quark inside a light meson
is found to decrease faster than $m_c^{-4}$ with increasing charm quark mass.
\end{abstract}

\maketitle

\section{Introduction}

The probability distributions of the momenta carried by light quarks and gluons inside a nucleon,
namely the parton distribution functions (PDFs), are the key nonperturbative ingredients to
unravel the nucleon internal structure.
In the past half century, the nucleon PDF has been determined with very high precision from numerous
high-energy collision experiments~\cite{Hou:2017khm}. Though the nucleon is viewed as a
baryon composed of three
light quarks in the context of naive quark model, it is generally believed that,
it must contain higher Fock components that entail heavy quark and anti-quark pair, {\it e.g.},
 $\vert uud c\bar{c}\rangle$, due to ubiquitous quantum fluctuation.
It has long been envisaged that the nucleon may have a non-negligible content of charm PDF,
usually dubbed {\it intrinsic} charm~\cite{Brodsky:1980pb,Brodsky:1981se,Brodsky:2015fna}.
Due to its nonperturbative nature, the intrinsic charm should be distinguished from the {\it extrinsic} charm,
which actually emerges from gluon splitting according to DGLAP evolution.
It has often been warned that the exact interpretation of intrinsic charm may suffer from some ambiguity.
For the notion of the intrinsic charm to make sense, the lifetime of an intrinsic $c\bar c$ pair inside a nucleon must be
much longer than the typical interaction time in the deep-inelastic scattering
processes~\cite{Blumlein:2015qcn}.

Recently, the \texttt{NNPDF} collaboration has released experimental evidence of existence of intrinsic charm in proton PDF
at a significance level of $3\sigma$~\cite{Ball:2022qks}.
They found that the very recent \texttt{LHCb} data~\cite{LHCb:2021stx} on $Z$ boson production associated with a charm jet
can be described very well only after including the intrinsic charm PDF in the analysis.
Previously, the \texttt{CTEQ-TEA} global analysis~\cite{Hou:2017khm} has placed an upper bound for the
average charm momentum fraction in a proton which is less than $2\%$ or $1.6\%$ at the renormalization scale $\mu=1.3 \; \mathrm{GeV}$.
The recent \texttt{NNPDF} article shows that average momentum fraction carried by the intrinsic charm is $0.62\%\pm 0.28\%$ at $\mu=1.65$ GeV~\cite{Ball:2022qks}.

It is very challenging to investigate the intrinsic charm PDF in a light hadron
directly from the first principle of QCD~\cite{Constantinou:2020hdm}. The Large Momentum Effective Theory (LaMET)~\cite{Ji:2013dva,Ji:2014gla,Ji:2020ect}
may have the bright potential to directly calculate the $x$-dependence of
intrinsic charm PDF on the lattice in the future.
However, at current stage, one has to resort to phenomenological models to parameterize the intrinsic charm PDF in a nucleon.
Two popular models are the meson cloud model (MCM)~\cite{Paiva:1996dd,Hobbs:2013bia,Melnitchouk:1997ig} and Brodsky-Hoyer-Peterson-Sakai (BHPS) model~\cite{Brodsky:1980pb,Brodsky:1981se}.
Unfortunately, it is not clear about the intimate connection between these two models and QCD.

Needless to say, it is highly desirable to understand the intrinsic charm PDF from a first-principle perspective.
Though formidably looking in realistic world, it is actually possible to achieve this goal in some toy models of QCD.
In this work, we attempt to investigate the intrinsic charm content of a light meson in the $1+1$ dimensional QCD in the large $N_c$ limit,
which was originally introduced by 't Hooft in 1974~\cite{tHooft:1974pnl}. Despite being a simple solvable model,
the 't Hooft model resembles the realistic QCD in several aspects, {\it e.g.}, color confinement, Regge trajectory, and chiral condensate.
A notable simplification in ${\rm QCD}_{2}$ is the lack of dynamic gluon. Once imposing light-cone gauge, the gluonic degree of freedom descends
simply to an interquark potential. Therefore, the charm quark PDF of a light meson in ${\rm QCD}_{2}$ has to be
 ``intrinsic" rather than ``extrinsic". The 't Hooft model thus may serve as an ideal theoretical laboratory to study the intrinsic charm PDF of a light hadron.
The aim of this work is to rigorously deduce the functional form of the intrinsic charm PDF inside a light meson in this toy model,
which starts at order-$1/N_c$.
To make a comparison, we also present the intrinsic charm PDF predicted by the light front two-dimensional counterparts of BHPS model and MCM.

The rest of this paper is distributed as follows. In Sec.~\ref{sec_thooft} we briefly review the Hamiltonian formalism of the
't Hooft model in the  $N_c\to\infty$ limit. In Sec.~\ref{sec_icpdf}
we extend the formalism to the next-to-leading order in $1/N_c$, and construct the functional form of the
intrinsic charm PDF with the aid of first-order quantum-mechanical perturbation theory.
In Sec.~\ref{sec_models}, we also give the explicit expressions of the intrinsic charm PDF within the two-dimensional versions of BHPS model and MCM.
We also discuss the relation between our rigorous result and the MCM result.
We devote Sec.~\ref{sec_num} to comprehensive numerical studies of the intrinsic charm PDF in a light meson which have been calculated by various approaches.
We also study how the first and second Mellin moments of the intrinsic charm PDF vary with the increasing charm mass.
Finally we summarize in Sec.~\ref{sec_sum}.

\section{A Brief Review of the Hamiltonian approach in 't Hooft Model\label{sec_thooft}}

In this section, we briefly review how to derive the 't Hooft equation using the light-front Hamiltonian method.
For more details, we refer the interested readers to Ref.~\cite{Jia:2018qee}.
The QCD Lagrangian in two spacetime dimensions reads
\begin{equation}
    \mathcal{L} =
    - \frac{1}{4}F^{a,\mu\nu}F^{a}_{\mu\nu}
    + \sum_{f}\overline\psi_{f}\left(i\slashed{D} - m_{f}\right)\psi_{f},
   \label{QCD:lagr}
\end{equation}
where $D_\mu= \partial_\mu-ig_s A_\mu^aT^a$ signifies the color covariant derivative
and $T^a$ denotes the generators of the $SU(N_c)$ group in the fundamental representation.
The gluon field strength tensor is defined as $F_{\mu\nu}^a \equiv \partial_\mu A_\nu^a-\partial_\nu A_\mu^a+g_sf^{abc}A_\mu^bA_\nu^c$.
$f$ denotes the flavor of quarks. In this work, we concentrate on the two-flavor case, where $f$ can be either
the up or the charm quark.
We use the chiral-Weyl representation for the Dirac $\gamma$ matrices:
\begin{equation}
    \gamma^0=\sigma_1,\quad \gamma^z=-i\sigma_2,\quad \gamma_5\equiv \gamma^0\gamma^z=\sigma_3,
\end{equation}
and the Dirac spinor field in this representation is
\begin{equation}
    \psi = 2^{-{\frac{1}{4}}}
    \left(
    \begin{array}{c}
        \psi_R \\ \psi_L
    \end{array}
    \right),
    \label{psidcp}
\end{equation}
where $R$, $L$ denote the right-handed and left-handed components, respectively.

The chiral limit and $N_c\to \infty$ limit do not generally commute.
In this work, we specify the 't Hooft model in the so-called ``weak-coupling" limit:
\begin{equation}
N_c\rightarrow\infty,\qquad \lambda \equiv \frac{g_s^2N_c}{4\pi}\;\,\text{fixed},\qquad m_{q}\gg g_s \sim {\frac{1}{\sqrt{N_c}}},
\label{eqn:lrgN_weak}
\end{equation}
where $\lambda$ of mass dimension two denotes the 't Hooft coupling constant.
We assume up quark to be light, so $m_u\le \sqrt{2\lambda}$; while charm quark is regarded as heavy,
hence $m_c\gg \sqrt{2\lambda}$.

It is convenient to adopt the light-cone coordinates  $x^\pm= x_{\mp}= (x^0\pm x^z)/\sqrt{2}$. Substituting \eqref{psidcp} into \eqref{QCD:lagr}, and imposing the light-cone gauge $A^{+, a} = 0$, one obtains
~\cite{tHooft:1974pnl}
\begin{align}
    \mathcal{L}
    =\frac{1}{2}
    \left(
        \partial_{-} A^{-,a}
    \right)^{2}
    + g_{s}\psi_{f,R}^{\dagger}A^{-,a}T^{a}\psi_{f,R}
    +\psi^{\dagger}_{f,R} i \partial_{+}\psi_{f,R}
    +\psi^{\dagger}_{f,L} i \partial_{-}\psi_{f,L}
    - \frac{m_f}{\sqrt{2}}
    \left(
        \psi^{\dagger}_{f,L}\psi_{f,R}+
        \psi^{\dagger}_{f,R}\psi_{f,L}
    \right) ,
\end{align}
where the flavor index $f$ is summed over $u$ and $c$.

In the light-cone gauge, $A^{-,a}$ and $\psi_{L}$ are no longer the dynamical variables.
From the equations of motion, they can be expressed in terms of the canonical variable $\psi_R$ (the ``good'' component):
\bsq
\begin{align}
    &\partial_{-}^{2}A^{-,a}-g_{s}\psi^{\dagger}_{R}T^{a}\psi_{R}=0,
    \\
    & i \partial_{-}\psi_{L}-\frac{m}{\sqrt{2}}\psi_{R}=0.
\end{align}
\esq
Substituting the solutions of these two equations into
the light-front Hamiltonian, we obtain
    \begin{align}
        H_{\mathrm{LF}} = P^- = \int_{x^{+}=\text{const.}}dx^{-}
        &\left[
            \frac{m^{2}}{2i}\psi^{\dagger}_{R}(x^{-})
            \int dy^{-} G^{(1)}(x^{-}-y^{-})\psi_{R}(y^{-})
        \right. \nonumber \\
              &
        \left.
            -\frac{g_{s}^{2}}{2}\
              \sum_{a} \psi^{\dagger}_{R}(x^{-})T^{a}\psi_{R}(x^{-})
              \int dy^{-} G^{(2)}(x^{-}-y^{-})
              \psi^{\dagger}_{R}(y^{-})T^{a}\psi_{R}(y^{-})
        \right],
    \end{align}
where $G^{(1)}$ and $G^{(2)}$ are the Green functions affiliated with the differential
operators $\partial_{-}$ and $\partial_{-}^{2}$:
\bsq
\begin{align}
        G^{(1)}(x^{-}-y^{-})&=i\int
        \frac{dk^{+}}{2\pi}
        \Theta{(|k^+|-\rho)}\frac{e^{-i k^{+}(x^{-}-y^{-})}}{k^{+}},
        \\
        G^{(2)}(x^{-}-y^{-})&=-\int
        \frac{dk^{+}}{2\pi}
        \Theta{(|k^+|-\rho)}
        \frac{e^{-ik^{+}(x^{-}-y^{-})}}{(k^{+})^{2}}.
    \end{align}
    \esq
Here $\rho$ is an artificial IR cutoff introduced to regularize the divergence caused by exchanging an instantaneous gluon.

One important feature of 't Hooft model is the color confinement. The isolated quarks and anti-quarks cannot
manifest themselves in physical spectrum.
It is the color-neutral quark-antiquark pair which can be created or annihilated in a physical process.
The technique of bosonization~\cite{Kikkawa:1980dc,Nakamura:1981zi,Rajeev:1994tr,Dhar:1994ib,Dhar:1994aw,Cavicchi:1993jh,Barbon:1994au,Itakura:1996bk}
turns out to be useful to diagonalize the light-front Hamiltonian.
One can define a set of color-singlet compound operators $M$,$B$ and $D$ from the quark/antiquark creation and annihilation operators:
\bsq
\begin{align}
    M^{\bar f_1f_2}\left(k^{+}, p^{+}\right)
    &=
    \frac{1}{\sqrt{N_c}} \sum_{c} d^{c,f_1}(k^{+})b^{c,f_2}(p^{+}),
    \\
    B^{f_1,f_2}\left(k^{+}, p^{+}\right)
    &=
    \sum_{c} b^{c,f_1\dagger}(k^{+})b^{c,f_2}(p^{+})
    \to
    \int^{\infty}_{0} \frac{dq^{+}}{2\pi}
    \sum_{f_{i}}
    M^{\dagger\bar f_{i}f_1}(q^{+}, k^{+})
    M^{\bar f_{i}f_2}(q^{+}, p^{+}),
    \\
    D^{\bar f_1,\bar f_2}\left(k^{+}, p^{+}\right)
    &=
    \sum_{c} d^{c,f_1\dagger}(k^{+})d^{c,f_2}(p^{+})
    \to
    \int^{\infty}_{0} \frac{dq^{+}}{2\pi}
    \sum_{f_{i}}
    M^{\dagger\bar f_1f_{i}}(k^{+}, q^{+})
    M^{\bar f_2f_{i}}(p^{+}, q^{+}),
\end{align}
\esq
where $c$ denotes the color index. The last equations reflect the color confinement assumption.

The commutation relation between $M$ and $M^\dagger$ is given by
\begin{equation}
    [M^{\bar f_1f_2}(k_1^{+},p_1^{+}), M^{\dagger\bar f_3f_4}(k_2^{+},p_2^{+})]
    =(2\pi)^{2}
    \delta_{f_1f_3}\delta_{f_2f_4}
    \delta(k_1^{+}-k_2^{+})\delta(p_1^{+}-p_2^{+})+\mathcal{O}\left(\frac{1}{N_c}\right),
\end{equation}
all other commutators among $M,B,D$ are at order $\mathcal{O}\left(1/N_c\right)$.
Since baryons become infinitely heavy and decouple in $N_c\to\infty$ limit, mesons are the only physical color-singlet states in this model.
One can diagonalize the light-front Hamiltonian by trading the compound operators $M$ and $M^\dagger$ for the mesonic annihilation and creation operators
$m_n$ and $m_n^\dagger$ ($n$ signifies the $n$-th excited meson). These two sets of operators are related by
the following relations:
{\bsq
\begin{equation}
    M^{\bar{f}_1f_2}((1-x)P^{+},xP^{+})
    = \sqrt{\frac{2\pi}{P^{+}}}
    \sum_{n=0}^{\infty}\varphi^{f_2\bar f_1}_{n}(x)m^{f_2\bar f_1}_{n}(P^{+}),
    \label{M to m}
\end{equation}
\begin{equation}
    m^{f_1 \bar{f}_2}_{n}(P^{+})
    = \sqrt{\frac{P^{+}}{2\pi}}
    \int^{1}_{0} dx
    \varphi_n^{f_1\bar f_2}(x)M^{\bar f_2 f_1}((1-x)P^{+}, xP^{+}),
    \label{m to M}
\end{equation}
\esq}
where the coefficient function $\varphi^{f_1\bar{f}_2}_n(x)$ is interpreted as the light-cone ('t Hooft) wave function of the $n$-th excited meson with the flavor
content $f_1\bar{f_2}$.

The meson annihilation and creation operators are assumed
to obey the standard commutation relation:
\begin{equation}
    \left[m^{f_i\bar f_j}_n(P^+_1), {m^{\dagger f_k\bar f_l}_r}(P^+_2)\right]
    =
    2\pi\delta_{f_if_k}\delta_{f_jf_l}\delta_{nr}\delta(P^+_1-P^+_2)
    + \mathcal{O}\left(\frac{1}{N_c}\right).
\label{m_commute}
\end{equation}

In order to have the desired commutation relation in \eqref{m_commute},
the 't Hooft wave functions must satisfy the following orthogonality and completeness conditions:
\begin{equation}
    \int^{1}_{0} dx\,
    \varphi^{f_1\bar f_2}_{n}(x)\varphi^{f_1\bar f_2}_{m}(x)
    =
\delta_{nm},
\label{orth and comp}
\end{equation}
\begin{equation}
    \sum_{n}
    \varphi^{f_1\bar f_2}_{n}(x)\varphi^{f_1\bar f_2}_{n}(y)
= \delta(x-y).
\end{equation}
At the leading order in $1/N_c$, the light-front Hamiltonian is simply a free Hamiltonian composed of all possible meson states:
\begin{equation}
    H_{\mathrm{LF}}= P^- = H_{\mathrm{vac}} +
    \sum_{n,f_1f_2}\int
    \frac{dP^{+}}{2\pi}P^{-}_{n,f_1f_2}
    m^{\dagger f_1\bar f_2}_{n}(P^{+})m^{f_1\bar f_2}_{n}(P^{+})
    + \mathcal{O}\left(\frac{1}{\sqrt{N_c}}\right).
    \label{eq:H_LF}
\end{equation}
The exact form of the vacuum energy $H_{\mathrm{vac}}$ can be found in \cite{Jia:2018qee}.  In order to reach such a
diagonalized Hamiltonian, the meson light-cone wave function must obey the celebrated 't Hooft equation~\cite{tHooft:1974pnl}:
\begin{equation}
    \left(\frac{m_1^2}{x}+\frac{m_2^2}{1-x}\right)
      \varphi_n^{f_1\bar f_2}\left(x\right) -2\lambda
    \dashint_0^1 dy\frac{\varphi_n^{f_1\bar f_2}\left(y\right)-\varphi_n^{f_1\bar f_2}\left(x\right)
    }{\left(x-y\right)^2} = \mu_{n,f_1,f_2}^2\varphi_n^{f_1\bar f_2}\left(x\right),
\end{equation}
where $m_{1}$, $m_2$ are the current quark masses affiliated with flavor $f_1$ and $f_2$, respectively, $\mu^{2}_{n,f_1f_2}$ is the squared meson mass.
The symbol $\dashint$ denotes the principal value (PV) prescription for an integral, defined as
\begin{equation}
    \dashint dy\, \frac{f(y)}{(x-y)^2}
    =
    \lim_{\epsilon \rightarrow 0^+}
    \int dy \, \Theta(|x-y|-\epsilon)
    {\frac{f(y)}{(x-y)^2}  } - {\frac{2 f(x)}{\epsilon}  }.
\end{equation}
Note that the IR regulator $\rho$ finally disappears from the LF Hamiltonian \eqref{eq:H_LF} as well as 't Hooft equation, as it should be.

\section{Intrinsic charm PDF of a light meson\label{sec_icpdf}}

Let us consider a light neutral meson composed of the $u$ and $\bar{u}$ quarks. For notational brevity, we simply call it $\pi$.
The intrinsic charm PDF of a pion follows the standard Collins-Soper definition~\cite{Collins:1981uw}:
\begin{equation}
    f_{c/\pi}(x) = \int \frac{dz^{-}}{4\pi}e^{-ixP^{+}z^{-}}\bra{\pi(P^+)}\overline
    c(z^{-})\gamma^{+}
    \mathcal{P} \left[\exp\left(-ig_s\int^{z^-}_0 d\eta^- A^+(\eta^-)\right)\right]
    c (0)\ket{\pi(P^+)}_{\mathrm{connected}},
\label{eq:pdfdef}
\end{equation}
where $P^+$ is the $+$-momentum of the pion, and $x$ is the $+$-momentum fraction carried by the charm quark with respect to the meson.
$c$ and $\overline{c}$ denote the charm quark fields, $\mathcal{P}[\cdots]$ denotes the gauge link which
ensures gauge invariance of the PDF. Since we have worked with the light-cone gauge $A^{a,+}=0$,
the gauge link can thus be simply dropped.

Employing the bosonization technique as mentioned in the preceding section,
the color-singlet non-local charm quark bilinear in \eqref{eq:pdfdef} can be expressed
in terms of the mesonic creation and annihilation operators:
    \begin{align*}
        \bar{c}(z^{-})\gamma^{+}
        c(0)
        =&
        c^\dagger_{R}(z^{-})c_{R}(0)
        \\
        =&
        \int \frac{dk^+_1dk^+_2}{2\pi} {N_{c}}\delta(k^+_1-k^+_2)e^{-ik^+_1z^{-}}
        \\&
        +\sum_{n}\int \frac{dk^+_1dk^+_2}{(4\pi)^{3/2}}\frac{\sqrt{N_{c}}}{\sqrt{k^+_1+k^+_2}}
        {e^{ik^+_1z^-}m^{\dagger c\bar{c}}_{n}}(k^+_1+k^+_2)
        {\varphi^{c\bar c}_{n}\left(\frac{k^+_1}{k^+_1\!+\!k^+_2}\right)}
        \\&
        +\sum_{n}\int \frac{dk^+_1dk^+_2}{(4\pi)^{3/2}}\frac{\sqrt{N_{c}}}{\sqrt{k^+_1+k^+_2}}
        {e^{-ik^+_1z^-}m^{c\bar{c}}_{n}}(k^+_1+k^+_2)
        {\varphi^{c\bar c}_{n}\left(\frac{k^+_2}{k^+_1\!+\!k^+_2}\right)}
        \\
        &+\sum_{f,n_1n_2}\int \frac{dk^+_1dk^+_2dq^+}{(2\pi)^{2}}e^{ik^+_1z^{-}}
        {m^{\dagger c\bar{f}}_{n_1}}(k^+_1+q^+)m^{c\bar{f}}_{n_2}(k^+_2+q^+)
        \frac{\varphi^{c\bar{f}}_{n_1}\left( \frac{k^+_1}{k^+_1+q^+} \right)}{\sqrt{k^+_1+q^+}}
        \frac{\varphi^{c\bar{f}}_{n_2}\left( \frac{k^+_2}{k^+_2+q^+} \right)}{\sqrt{k^+_2+q^+}}\\
        &-\sum_{f,n_1n_2}\int \frac{dk^+_1dk^+_2dq^+}{(2\pi)^{2}}e^{-ik^+_1z^{-}}
        {m^{\dagger f\bar{c}}_{n_1}}(k^+_2+q^+)m^{f\bar{c}}_{n_2}(k^+_1+q^+)
        \frac{\varphi^{f\bar{c}}_{n_1}\left( \frac{q^+}{k^+_2+q^+} \right)}{\sqrt{k^+_2+q^+}}
        \frac{\varphi^{f\bar{c}}_{n_2}\left( \frac{q^+}{k^+_1+q^+} \right)}{\sqrt{k^+_1+q^+}}.
        \addnumber
        \label{eq:bilinear_exp}
    \end{align*}
The ${\cal O}(N_c)$ term contributes to the disconnected part, thus can be dropped.
The ${\cal O}(\sqrt{N_c})$  terms only contain a single meson creation or annihilation operator,
which also make vanishing contribution when sandwiched between two $\pi$ states.
Only last two terms of ${\cal O}(N_c^0)$ yield non-vanishing contribution,
which represent the charmed meson sector and anti-charmed meson sector, respectively.

Next we turn to the higher Fock component inside a physical $\pi$ state.
In the $N_c\to \infty$ limit, the $\pi$ only contains the valence constituents $u\bar u$.
In order to nail down its intrinsic charm content, one has to expand the ${\rm QCD}_2$ light-front Hamiltonian to next-leading order in $1/N_c$.
Let us split the full Hamiltonian into $H_{{\text{LF}}}= H_{{\text{LF}},0}+V$, where the free mesonic Hamiltonian
$H_{{\text{LF}},0}$ is given in \eqref{eq:H_LF},
and the $V$ term encapsulates all possible ${\cal O}(1/\sqrt{N_c})$ three-meson interactions.
Invoking the first-order quantum-mechanical perturbation theory, the physical pion state can be expressed as
\begin{equation}
\ket{\msn'}
\approx \ket{\msn} + \frac{1}{P^--H_{\mathrm{LF},0}+i\epsilon}V \ket{\msn}.
\label{lip-sch equation}
\end{equation}
$\ket{\pi'}$ denotes the eigenstate of the full Hamiltonian, and $\ket{\pi}$ signifies the eigenstate of $H_{{\text{LF}},0}$,
which can be generated by
\begin{equation}
    \ket{\pi_n(P^+)} = \sqrt{2P^+}m_n^{\dagger u\bar u}(P^+)\ket{0},
\label{eq:pi_n:m}
\end{equation}
here $n$ denotes the principle quantum number.

It is well-known that the  ${\cal O}(1/\sqrt{N_c})$ piece of the interaction potential $V$ is governed by three-meson coupling \cite{Callan:1975ps}.
To our concern, the most relevant parts in $V$  are those coupling $\pi$ with all possible charmed mesons and anti-charmed mesons:
\begin{equation}
    V_{\rm charm}=\mathcal{V}+\overline{\mathcal{V}}+\mathrm{h.c.},
    \label{Vc}
\end{equation}
where
\bsq
\begin{align*}
        \mathcal{V} =
        \frac{-\lambda}{(2\pi)^{ \frac{3}{2} }\sqrt{N_{c}}}
        &
        \sum_{n_1n_2n_3}
        \int_0^\infty dq^+ dk^+_1dk^+_2dk^+_3dk^+_4
        \delta(k^+_1-k^+_2+k^+_3+k^+_4)
        {m^{\dagger c\bar{u}}_{n_1}}(k^+_1+q^+)
        m^{u\bar{u}}_{n_2}(k^+_2+q^+)
        {m^{\dagger u\bar{c}}_{n_3}}(k^+_3+k^+_4) \\
           &\ \ \ \ \ \ \ \ \
           \times\frac{1}{(k^+_3-k^+_2)^{2}}
           \frac{\varphi^{c\bar{u}}_{n_1}\left(\frac{k^+_1}{k^+_1+q^+}\right)}{\sqrt{k^+_1+q^+}}
           \frac{\varphi^{u\bar{u}}_{n_2}\left(\frac{k^+_2}{k^+_2+q^+}\right)}{\sqrt{k^+_2+q^+}}
           \frac{\varphi^{u\bar{c}}_{n_3}\left(\frac{k^+_3}{k^+_3+k^+_4}\right)}{\sqrt{k^+_3+k^+_4}}
        \addnumber
        \label{V},
    \end{align*}
    \begin{align*}
        \overline{\mathcal{V}} =
        \frac{\lambda}{(2\pi)^{ \frac{3}{2} }\sqrt{N_{c}}}
        &
        \sum_{n_1n_2n_3}
        \int_0^\infty dq^+ dk^+_1dk^+_2dk^+_3dk^+_4
        \delta(k^+_1-k^+_2+k^+_3+k^+_4)
        {m^{\dagger u\bar{c}}_{n_1}}(k^+_1+q^+)
        m^{u\bar{u}}_{n_2}(k^+_2+q^+)
        {m^{\dagger c\bar{u}}_{n_3}}(k^+_3+k^+_4) \\
           &\ \ \ \ \ \ \ \ \ \
           \times\frac{1}{(k^+_3-k^+_2)^{2}}
           \frac{\varphi^{u\bar{c}}_{n_1}\left(\frac{q^+}{k^+_1+q^+}\right)}{\sqrt{k^+_1+q^+}}
           \frac{\varphi^{u\bar{u}}_{n_2}\left(\frac{q^+}{k^+_2+q^+}\right)}{\sqrt{k^+_2+q^+}}
           \frac{\varphi^{c\bar{u}}_{n_3}\left(\frac{k^+_4}{k^+_3+k_4}\right)}{\sqrt{k^+_3+k^+_4}}
        \addnumber
        \label{Vbar}.
    \end{align*}
\label{VVbar}
\esq
Note that $V_{\rm charm}$  is indeed of order-$1/\sqrt{N_c}$. Obviously,
the interaction potential $V_{\rm charm}$ can induce transitions from $\pi$ into a $D\overline{D}$ pair,
with $D$ and $\overline{D}$ generically referring to all possible excited charmed and anti-charmed mesons.

To proceed, let us insert a complete set of hadronic states in the left of $V_{\rm charm}$ in Eq.\eqref{lip-sch equation}.
Clearly, only those intermediate states composed of free $D\overline{D}$ pairs can survive in the sum.
Eq.~\eqref{lip-sch equation} can then be recast as
\begin{align}
\ket{\pi_n'(P^+)} &\approx \ket{\pi_n(P^+)} + \sum_{n_in_j} \int^{\infty}_{0}
      \frac{dk^+_idk^+_j}{(2\pi)^{2}2k^+_i 2k^+_j}
      \widetilde{T}_{n,n_i,n_j}(k^+_i,k^+_j)
      \ket{D_{n_i}(k^+_i)\overline{D}_{n_j}(k_j^+)},
\label{physical:pion:expansion}
\end{align}
where the completed charmed hadronic states arising from the first-order perturbation are defined by
    \begin{equation}
        \ket{D_{n_i}(k^+_i),\overline D_{n_j}(k^+_j)}=2\sqrt{k^+_ik^+_j}
        m^{\dagger c\bar{u}}_{n_i}(k^+_i)
        m^{\dagger u\bar{c}}_{n_j}(k^+_j) \ket{0},
        \label{D state}
    \end{equation}

For the sake of generality, here we consider the intrinsic charm content of the $n$-th excited pion state (denoted by $\pi_n$),
rather than only consider the ground-state $\pi$.
The $\widetilde{T}$ function in \eqref{physical:pion:expansion} is defined as
 \begin{equation}
     \widetilde{T}_{n,n_i,n_j}(k^+_i, k^+_j)
     \equiv
     \braket{D_{n_i}(k_i^+)\overline{D}_{n_j}(k_j^+) | \pi_{n}'(P^+)}
     \approx
     \bra{D_{n_i}(k_i^+)\overline{D}_{n_j}(k_j^+)}V_{\rm charm}\ket{\pi_{n}(P^+)}
   \left(
         \frac{\mu^{2}_{D_{n_i}}}{2k^+_i}+
         \frac{\mu^{2}_{\overline D_{n_j}}}{2k^+_j}-
         \frac{\mu^{2}_{\pi_{n}}}{2P^+}
     \right)^{-1}.
 \label{Ttilde:probability:ampl}
 \end{equation}
This function has a clear physical interpretation, which characterizes the probability amplitude of finding a $\ket{D\overline D}$ state with certain quantum number
in a physical $\pi_n$~\footnote{Note we have dropped the $i\epsilon$ term in the energy denominator in \eqref{Ttilde:probability:ampl}, because the energy denominator has always positive
sign due to $P^+\ge k_i^+,k_j^+ \ge 0$ and $\mu_D, \mu_{\overline D} \gg \mu_\pi$, if we do not consider the excessively highly-excited pion.}.

\begin{figure}[htbp]
\centering
\includegraphics[width=0.45\textwidth]{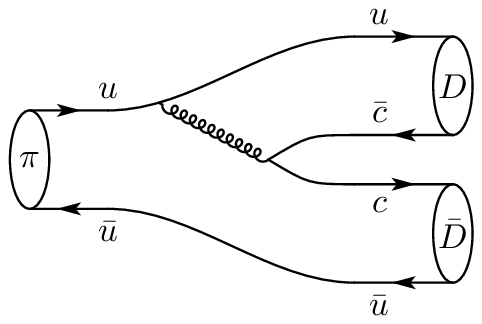}
\includegraphics[width=0.45\textwidth]{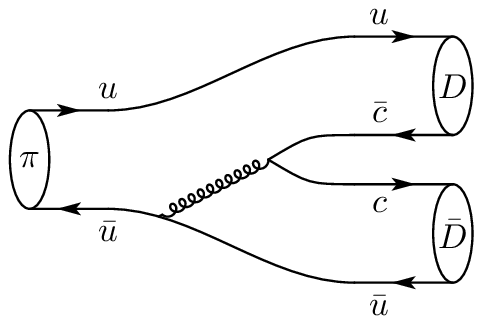}
\caption{Schematic diagrams illustrating pion transitioning into a $D\overline{D}$ pair.}
\label{fig:meson_vrtx}
\end{figure}

The matrix element in \eqref{Ttilde:probability:ampl} can be further expressed as
\begin{equation}
    \bra{D_{n_1}(x_1P^+) \overline{D}_{n_2}(x_2P^+)} V_{\rm charm} \ket{\pi_n(P^+)} = {2\pi\over P^+}\delta(x_1+x_2-1)\varGamma_{n,n_1,n_2}(x_1,x_2),
\label{gammadef}
\end{equation}
where $x_i=k^+_i/P^+$ ($i=1,2$) is the $+$-momentum fraction of $D,\overline{D}$ with respect to $\pi_n$, and this matrix element vanishes unless the
light-cone momentum conservation is satisfied.
The transition vertex function $\varGamma$  has been first given by Callan, Coote and Gross long ago~\cite{Callan:1975ps}, whose
explicit form reads
\begin{align}
    \varGamma_{n,n_1,n_2}\left(x_1, x_2\right)
    =&
   4\lambda \sqrt{\frac{\pi}{N_{c}}}
    \left[
        \int^1_{x_1}dy_1\int^{x_1}_0dy_2\frac{1}{(y_2-y_1)^{2}}
        \varphi^{u\bar{u}}_{n}\left( 1-y_1\right)
        \varphi^{c\bar{u}}_{n_1}\left(1-\frac{y_2}{x_1}\right)
        \varphi^{u\bar{c}}_{n_2}\left(\frac{1-y_1}{x_2}\right)
    \right.
\nonumber
\\
&\qquad\quad
\left.
-\int^1_{x_2}dy_1\int^{x_2}_0dy_2\frac{1}{(y_2-y_1)^{2}}
        \varphi^{u\bar{u}}_{n}\left( y_1\right)
        \varphi^{u\bar{c}}_{n_2}\left(\frac{y_2}{x_2}\right)
        \varphi^{c\bar{u}}_{n_1}\left(\frac{y_1-x_2}{x_1}\right)
\right].
\label{vertex}
\end{align}
In Fig.~\ref{fig:meson_vrtx} we present some schematic diagrams depicting the triple-meson vertex $\varGamma$.

It is reassuring to see that the vertex function $\varGamma$ is indeed of order $N_c^{-1/2}$.
At first sight, one may worry that the $\varGamma$ may become divergent
when the integration variables approach the boundary, {\it i.e.}, $y_1,y_2 \to x_1$ or $x_2$.
A careful look reveals that near the boundary, both terms in the integrand are simultaneously
approaching  $\varphi^{u\bar u}_{n}(x_2) \varphi^{c\bar u}_{n_1}(0)\varphi^{u\bar c}_{n_2}(1) $,
therefore the potential IR divergences cancel, so the vertex function $\varGamma$
is warranted to be IR finite.

\begin{figure}[htbp]
\begin{minipage}{\textwidth}
\centering
\raisebox{-0.5\height}{\includegraphics[width=0.45\textwidth]{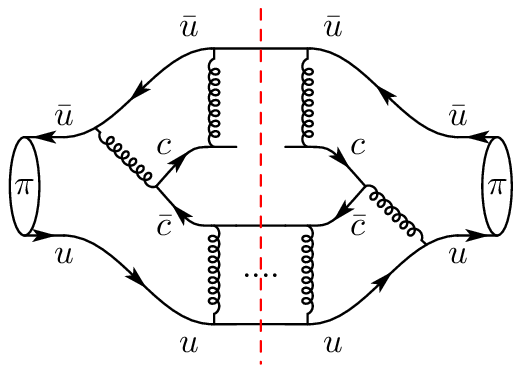}}
\raisebox{-0.5\height}{\includegraphics[width=0.45\textwidth]{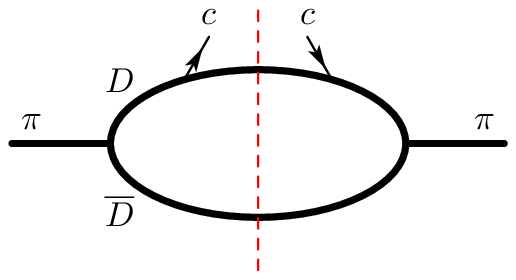}}
\end{minipage}
\caption{Schematic figures for our rigorous result of intrinsic charm PDF of $\pi$ (left) and for meson cloud model (right).}
\label{fig:factrize}
\end{figure}

Substituting Eqs.~\eqref{eq:bilinear_exp} and \eqref{physical:pion:expansion} into the PDF definition in \eqref{eq:pdfdef},
and repeatedly using the commutation relation in \eqref{m_commute}, we can express the intrinsic charm PDF of the $\pi_n$ as
\begin{equation}
    f_{c/\pi_n}(x)
    \!=\!\!\sum_{n_1n_2n_3n_4} \!\!
    \int\!\! \frac{dx_1 dx_2 dx_3 dx_4}{16(2\pi)^{4}x_1 x_2 x_3 x_4}
    \widetilde{T}_{n,n_1,n_2}(x_1P^+, x_2P^+)\,
        \mathcal{H}^{n_1,n_2}_{n_3,n_4}(x_1,x_2,x_3,x_4,x)\,
     \widetilde{T}^*_{n,n_3,n_4}(x_3P^+,x_4P^+),
\label{eq:icpdf_fact}
\end{equation}
with
\begin{align}
\label{H:n1:n2:n3:n4}
    \mathcal{H}^{n_1,n_2}_{n_3,n_4}(x_1,x_2,x_3,x_4,x)
    &\equiv
        \int \frac{dz^{-}}{4\pi} e^{-ixP^{+}z^{-}}\!\!
            \bra{D_{n_3}(x_3P^+)\overline{D}_{n_4}(x_4P^+)}
            c^\dagger_{R}(z^{-}) c_{R}(0)
            \ket{D_{n_1}(x_1P^+)\overline{D}_{n_2}(x_2P^+)}
 \nonumber   \\
    &= 4 \pi
        \left[
        \delta_{n_2 n_4}
        x_2\delta(x_4-x_2)
        \theta(x)
        \theta(x_3-x)
        \varphi^{c\bar{u}}_{n_3}\left(\frac{x}{x_3}\right)
        \varphi^{c\bar{u}}_{n_1}\left(\frac{x+x_1-x_3}{x_1}\right)
    \right.
    \\
    &\ \ \ \ \ \ \ \
    \left.
        -\delta_{n_1 n_3}
        x_1\delta(x_3-x_1)
        \theta(-x)
        \theta(x_2+x)
        \varphi^{u\bar{c}}_{n_2}\left(\frac{x+x_2}{x_2}\right)
        \varphi^{u\bar{c}}_{n_4}\left(\frac{x+x_2}{x_4}\right)
    \right],
\nonumber
\end{align}
in which $\theta(x)$ and $\theta(-x)$ terms represent the charm and anti-charm sectors, respectively.

Substituting the definition of $\widetilde{T}$ in \eqref{Ttilde:probability:ampl} into \eqref{eq:icpdf_fact}, and
we finally arrive at a compact form of the intrinsic charm PDF of the $\pi_n$:
\begin{align}
\label{eq:icpdfexp}
   f_{c/\pi_n} (x)
    =&\sum_{n_1,n_2,n_3,n_4}
    \int dx_1
    \frac{
        \varGamma_{n,n_1, n_2}(x_1)\varGamma^*_{n,n_3, n_4}(x_1)
    }{16\pi  x_{1}(1-x_1)}
    \left(
        \frac{\mu^{2}_{D_{n_1}}}{2x_1}+
        \frac{\mu^{2}_{\overline D_{n_2}}}{2(1-x_1)}-
        \frac{\mu^{2}_{\pi_{n}}}{2}
    \right)^{-1}
    \left(
        \frac{\mu^{2}_{D_{n_3}}}{2x_1}+
        \frac{\mu^{2}_{\overline D_{n_4}}}{2(1-x_1)}-
        \frac{\mu^{2}_{\pi_{n}}}{2}
    \right)^{-1}
    \\&
    \times\!
    \left[
        \frac{\theta(x)\theta(x_1\!-\!x)}{x_1}
        \delta_{n_2 n_4}
        {
            \varphi^{c\bar{u}}_{n_3}\!\left(\frac{x}{x_1}\right)
            \varphi^{c\bar{u}}_{n_1}\!\left(\frac{x}{x_1}\right)
        }
        -
        \frac{\theta(-x)\theta(x\!-\!x_1\!+\!1)}{1-x_1}
        \delta_{n_1 n_3}
       {
            \varphi^{u\bar{c}}_{n_2}\!\left(1\!+\!\frac{x}{1\!-\!x_1}\right)
            \varphi^{u\bar{c}}_{n_4}\!\left(1\!+\!\frac{x}{1\!-\!x_1}\right)
        }
    \right].
    \nonumber
\end{align}

Equation~\eqref{eq:icpdfexp} is the main result of this work, which represents the rigorous expression for
the intrinsic charm PDF of a light meson in the 't Hooft model.
A schematic Feynman diagram to visualize this formula
is shown in the left figure in Fig.~\ref{fig:factrize}.
The most important message is that, there is an infinite tower of charmed mesons and anti-charmed mesons that manifest as the higher Fock components of a light meson
and contribute to the intrinsic charm PDF.
Note the principle quantum numbers $n_1$, $n_2$, $n_3$ and $n_4$ in the sum over (anti-)charmed mesons are all
independent.
It is worth mentioning that, if we only keep the diagonal terms in the sum,  {\it i.e.},
taking $n_1=n_3$ and $n_2=n_4$ simultaneously, Eq.~\eqref{eq:icpdfexp} reduces to the prediction of the intrinsic charm PDF from the meson
cloud model. We will discuss the derivation of the intrinsic charm PDF in MCM  in details in next section.

\section{The BHPS model and meson cloud model in ${\rm QCD}_2$ \label{sec_models}}

The BHPS model is a very simple and intuitive model to parametrize the intrinsic charm PDF of a light hadron.
The key assumption is that the four-quark Fock component $\vert u\bar{u} c\bar{c}\rangle$ in $\pi$ can be treated as a free four-body state~\cite{Brodsky:1980pb,Brodsky:1981se}.
The intrinsic charm PDF can be then approximated from the transition probability of $\ket{\pi} \to \ket{u\bar u c \bar c}$ by
the first-order light-front perturbation theory:
\begin{equation}
    \frac{d \mathrm{Prob.}}{dx_u dx_{\bar{u}} dx_c dx_{\bar{c}}} \propto
    \delta(1-x_u-x_{\bar{u}} - x_c - x_{\bar{c}})
    \left(
        m^2_\pi
        - \frac{m_u^2}{x_u}
        - \frac{m^2_{\bar{u}}}{x_{\bar u}}
        - \frac{m^2_c}{x_c}
        - \frac{m_{\bar{c}}^2}{x_{\bar c}}
    \right)^{-2},
\label{eq:bhps_exact}
\end{equation}
where $x_i$ indicates the $+$-momentum fraction carried by each parton.

In the heavy quark limit $m_c\gg m_u, m_\pi$, one can drop small quantities in the energy denominator, and
\eqref{eq:bhps_exact} reduces to
\begin{equation}
    \frac{d \mathrm{Prob.}}{dx_u dx_{\bar{u}} dx_c dx_{\bar{c}}}  \propto
    \delta(1-x_u-x_{\bar u} - x_c - x_{\bar c})
    \left(\frac{x_c x_{\bar c}}{x_c+x_{\bar c}}\right)^2.
\label{eq:bhps:approx}
\end{equation}

Integrating \eqref{eq:bhps:approx} over $x_u$, $x_{\bar u}$, $x_{\bar c}$, one then arrives at the intrinsic charm PDF predicted
by the BHPS model:
\begin{equation}
\label{eq:bhps_exp}
f_c(x)= Ax^{2} \left[ \frac{1}{2}\left(1+4x-5x^{2}\right) +x\left(2+x\right)\ln x \right],
\end{equation}
where $A$ is an unknown normalization constant which can not be determined within the BHPS model itself.
There are three popular variants of the BHPS model. BHPS1, BHPS2 determine the parameter $A$ through different
global fit recipes,
and the BHPS3 model takes numerical integration directly following \eqref{eq:bhps_exact}~\cite{Hou:2017khm}.
Inspired by the ansatz of the BHPS model, Pumplin parameterized the intrinsic charm PDF of a proton
using a five-quark model including quarks' transverse motion~\cite{Pumplin:2005yf}.

Another influential model is the meson cloud model that assumes the proton has non-negligible five-quark Fock component composed of a
charmed baryon and a charmed meson due to
inevitable quantum fluctuation~\cite{Paiva:1996dd,Hobbs:2013bia,Melnitchouk:1997ig}.
In the context of current work, the relevant quantum fluctuation inside $\pi$ is the higher Fock component composed of the
 charmed and anti-charmed mesons. According to the spirit of MCM,
 the intrinsic charm PDF of the $\pi_n$ is expressed as the transition probability of $\pi_n \to D_{n_1}\overline{D}_{n_2}$
 convoluted with the valence charm PDF inside the charmed meson $D_{n_1}$:
\begin{equation}
    f_{c/\pi_n}(x)
    =
    \sum_{n_1,n_2}
    \int^1_0 dy \mathcal{F}_{n,n_1,n_2}(y)
    \int^1_0 d\eta f_{c/D_{n_1}}(\eta)\delta(x-\eta y)
    =
    \sum_{n_1,n_2}
    \int^1_x \frac{dy}{y}
    \mathcal{F}_{n,n_1,n_2}(y)f_{c/{D_{n_1}}}\left(\frac{x}{y}\right)
    \label{eq:mcm}
\end{equation}
where $\mathcal{F}_{n,n_1,n_2}(y)$ denotes the transition probability of
$\pi_n$ with $+$-momentum $P^{+}$ transitioning into a charmed meson $D_{n_1}$ that carries the $+$-momentum $yP^{+}$,
and $f_{c/D_{n_i}}(x)$ denotes the valence charm PDF of the charmed meson $D_{n_i}$.

Let us first consider the transition probability factor $\mathcal{F}$ accompanied with process  $\pi_n\to D_{n_1}(x_1),\overline{D}_{n_2}$:
\begin{equation}
\mathcal{F}_{n,n_1n_2}(x_1)dx_1 =
        \frac{1}{\widetilde{V}} \frac{1}{2P^+}
        \frac{P^{+}dx_1}{(2\pi) (2x_1P^+)}
        \int \frac{P^{+}dx_2}{(2\pi) (2x_2P^+)}
        \left|\braket{ D_{n_1}(x_1P^+)\overline{D}_{n_2}(x_2P^+)|\pi'_n(P^+)}\right|^{2}.
        \label{fnnndef}
\end{equation}
Note that the inner product in \eqref{fnnndef} is exactly the $\widetilde{T}$ function defined in
\eqref{Ttilde:probability:ampl}, which characterizes the probability amplitude of finding a specific $D\overline{D}$ state inside the $\pi_n$.
Since we do not care about the anti-charmed meson $\overline{D}_{n_2}$, a phase space integration should be assigned to the $+$-momentum fraction
$x_2$ carried by the $\overline{D}_{n_2}$ meson. However, as indicated in \eqref{gammadef}, $+$-momentum conservation demands that the
$\widetilde{T}$ function contains a $\delta$-function $\delta(1-x_1-x_2)$. Therefore the integration over $x_2$ becomes trivial.
Interestingly, the factor $1/\widetilde{V}$ can help eliminate the ill-defined $\delta(0)$ arising from
squaring the $\widetilde{T}$ function, since the finite volume $\widetilde V $
can be identified with $2\pi\delta_V(0\times P^{+})$ in the box quantization.

Substituting \eqref{Ttilde:probability:ampl} and \eqref{gammadef} into \eqref{fnnndef}, it is straightforward to obtain
\begin{align}
\label{exact:F:function}
        \mathcal{F}_{n,n_1n_2}(x_1)
        =&
            \frac{1}{16\pi}
        \frac{1}{x_1(1-x_1)}
        \frac{
            \left|\varGamma_{n,n_1n_2}(x_1)\right|^{2}
        }{
            \left(
                \frac{\mu^{2}_{D_{n_1}}}{2x_1}+
                \frac{\mu^{2}_{\bar{D}_{n_2}}}{2(1-x_1)}-
                \frac{\mu^{2}_{\pi_{n}}}{2}
            \right)^{2}
        }.
    \end{align}

In passing, we emphasize that our rigorous result for intrinsic charm PDF in \eqref{eq:icpdfexp}
automatically includes the probability of finding anti-charm, since it satisfies the relation $f_{\bar{c}/\pi}(x)= - f_{c/\pi}(-x)$ due to charge conjugation
symmetry inherent in the PDF definition \eqref{eq:pdfdef} for a neutral $\pi$ meson.
In order to make an intimate comparison between the meson cloud model prediction and our rigorous result in \eqref{eq:icpdfexp},
the anti-charm sector should also be explicitly added in the MCM, hence we generalize $\eqref{eq:mcm}$ as
\begin{align}
    f_{c/\pi_n}(x)
    &=
    \sum_{n_1,n_2}
    \int^1_0 dy \left(
        \mathcal{F}_{n,n_1,n_2}(y)
        \int^1_0 d\eta f_{c/D_{n_1}}(\eta)\delta(x-\eta y)
        -
        \mathcal{F}_{n,n_1,n_2}(1-y)
        \int^1_0 d\eta f_{\bar c/\overline D_{n_2}}(\eta)\delta(-x-\eta y)
        \right)
   \nonumber \\
    &=
    \sum_{n_1,n_2}
    \int^1_0 dy
    \mathcal{F}_{n,n_1,n_2}(y)
    \left(
    \theta(x)\theta{(y-x)}
    \frac{f_{c/D_{n_1}}\left(\frac{x}{y}\right)}{y}
    -
    \theta(-x)\theta{(x-y+1)}
   \frac{f_{\bar c/\overline D_{n_2}}\left(-\frac{x}{1-y}\right)}{1-y}
    \right)
\label{MCMphi}
  \\
    &=
    \sum_{n_1,n_2}
    \int^1_0 dy
    \mathcal{F}_D(y)
    \left(
    \theta(x)\theta{(y-x)}
    \frac{1}{y}
    \left[\varphi^{c\bar u}_{n_1}\left(\frac{x}{y}\right)\right]^2
    -
    \theta(-x)\theta{(x-y+1)}
    \frac{1}{1-y}
    \left[\varphi^{u\bar c}_{n_2}\left(1+\frac{x}{1-y}\right)\right]^2
    \right).
    \nonumber
\end{align}
In \eqref{MCMphi} we have made use of the knowledge that the valence charm PDF inside a $D$ meson
is simply the square of the corresponding 't Hooft wave function:
\bsq
\begin{equation}
    f_{c/D_n}(x) = \left[\varphi^{c\bar u}_n(x) \right]^2,
\end{equation}
\begin{equation}
    f_{\bar c/\overline D_n}(x) = f_{u/\overline D_n}(1-x) = \left[\varphi^{u\bar c}_n(1-x)\right]^2.
\end{equation}
\esq

Plugging \eqref{exact:F:function} in  \eqref{MCMphi}, we obtain the final prediction of the intrinsic charm PDF
given by MCM
\begin{align}
        f_{c/\pi_n} (x)
        =&
        \sum_{n_1,n_2}
        \int dx_1
        \frac{|\varGamma_{n,n_1n_2}(x_1)|^2}{16\pi x_{1}(1-x_1)}
        \left(
            \frac{\mu^{2}_{D_{n_1}}}{2x_1}+
            \frac{\mu^{2}_{\overline D_{n_2}}}{2(1-x_1)}-
            \frac{\mu^{2}_{\pi_{n}}}{2}
        \right)^{-2}
       \nonumber
       \\
       &  \times
        \left(
            \theta(x)\theta(x_1-x)
            \frac{
                \left[\varphi^{c\bar{u}}_{n_1}\left(\frac{x}{x_1}\right)\right]^2
            }{x_1}
            -
            \theta(-x)\theta(x-x_1+1)
            \frac{
                \left[\varphi^{u\bar{c}}_{n_2}\left(1+\frac{x}{1-x_1}\right)\right]^2
            }{1-x_1}
        \right).
        \addnumber
\label{result 2}
    \end{align}
A schematic Feynman diagram to picturise the MCM
is shown in the right figure in Fig.~\ref{fig:factrize}.

It is amazing that the MCM prediction of the intrinsic charm PDF looks quite similar to our rigorous result
in \eqref{eq:icpdfexp}, except the latter does not enforce the diagonal condition $n_1=n_3$ and $n_2=n_4$,
and the `interference' terms with $n_1\neq n_3$ or $n_2\neq n_4$ in (\ref{eq:icpdfexp})
do make important contribution.

It is interesting to note that, because of the orthogonality relation of the
't Hooft wave functions as in \eqref{orth and comp},
the `interference' terms do not contribute to the first Mellin moment of the intrinsic charm PDF.
Of course, they will affect the shape of the intrinsic charm PDF and the average charm momentum fraction.

\section{Numerical Results\label{sec_num}}

In this section, we present the numerical results of intrinsic charm PDF in a fictitious pion meson. In the large $N_c$ limit, we set the mass scale following the ansatz in Ref.~\cite{Burkardt:2000uu},
by choosing the value of the 't Hooft coupling $\sqrt{2\lambda}=340\,\mathrm{MeV}$ in correspondence to the value of string tension in the realistic QCD. 
 To save calculational labor, we deliberately choose the up quark mass $m_u=0.749\sqrt{2\lambda}$, 
which is equal to the strange quark mass determined in \cite{Jia:2017uul}. We also studied intrinsic charm content inside a pion with different values of charm quark mass. The charm mass is varied from $m_c=4.19\sqrt{2\lambda}$ to $m_c = 3 m_b$ with $m_b=13.66\sqrt{2\lambda}$. 
For the details of setting masses of different quark flavors, we refer the interested readers to Ref.~\cite{Jia:2017uul,Jia:2018mqi}.

The light-cone wave functions of the $u\bar{u}$, $u\bar{c}$ and $c\bar{u}$ state sare obtained by solving the 't Hooft equation by means of the Brower-Spencer-Weis (BSW) method~\cite{Brower:1978wm}. 
We use 120 BSW bases for the cases $m_c < 13.66\sqrt{2\lambda}$, 192 BSW bases for $13.66\sqrt{2\lambda}\leq m_c\leq 27.32\sqrt{2\lambda}$ and 264 BSW bases for $m_c>27.32\sqrt{2\lambda}$.

We calculate the intrinsic charm PDF according to our rigorous expression \eqref{eq:icpdfexp}, as well as the predictions given by MCM and BHPS model. 
To make a fair comparison, we normalize the results of BHPS model to have the equal first Mellins moment as that of the rigorous result and the MCM.  
In our analysis we also include a naive meson cloud model, which only includes the ground state in the sum in \eqref{result 2}.

The intrinsic charm PDF from our rigorous calculation in \eqref{eq:icpdfexp} and MCM~\eqref{result 2} involve a sum over all possible intermediate meson states. 
We impose a truncation $n_{1,2,3,4}\leq N$ to facilitate the summation. Due to the limitation of our computing resources, the maximum value of $N$ is set to 
$N_{\mathrm{max}}=60$.  The convergence criteria is set by searching for the lowest $N=N_{0}$ that satisfies
\begin{equation}
\frac{\displaystyle \int_0^1 dx\, \left[f_c^{(N)}(x)-f_c^{(N_{\text{max}})}(x)\right]^2}
{\displaystyle \int_0^1 dx\, \left[\frac{f_c^{(N)}(x)+f_c^{(N_{\text{max}})}(x)}{2}\right]^2}\leq 0.01, 
\end{equation}
where $f_{c}^{(N)}(x)$ denotes the intrinsic charm PDF in Eq.~\eqref{eq:icpdfexp} with summation truncated at $N$.
For instance, the intrinsic charm PDF of the first excited pion converges at $N_0=34$ when $m_c=4.19\sqrt{2\lambda}$ and $N_0=48$ when $m_c=13.66\sqrt{2\lambda}$. The intrinsic PDF from MCM shows a better converge tendency, thus we take $N=N_\mathrm{max}$ as the final results for MCM.

\begin{figure}[htbp]
\includegraphics[width=0.9\textwidth]{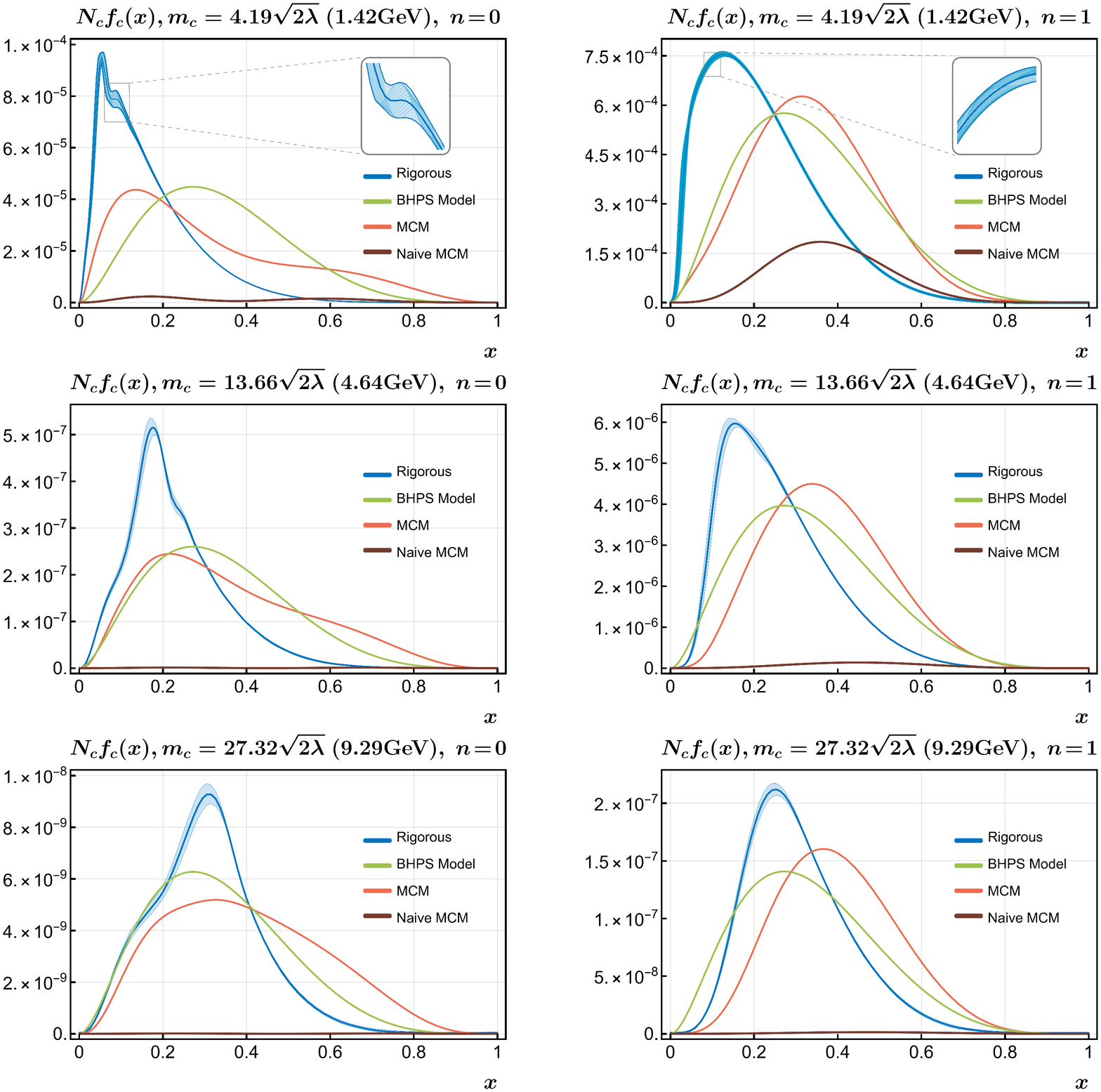}
\caption{Intrinsic charm PDF in pion from our rigorous analysis, MCM, Naive MCM and BHPS models. 
We show the results with some representative values of charm quark masses. 
The left and right columns show the results of a ground state and first excited state $\pi$, respectively. 
The thin dark blue curves in the windows correspond to each $N$ lying between $N_0$ and $60$.}
\label{fig:pdffig}
\end{figure}

The contribution from high excited states in Eq.~\eqref{eq:icpdfexp} only affects the microscopic texture of intrinsic charm PDF. 
We treat the choices of $N$ as source of systematic uncertainties in our calculation. 
To be more specific, we plot the envelopes of curves corresponding to quark PDF with $N_0\leq N \leq N_\mathrm{max}$ and use the upper and lower envelope as upper and lower bound correspondingly. 
The central value is given as the average of the upper and lower bound. To demonstrate how to determine the upper and lower bound, we magnify part of the curve corresponding to our rigorous results 
in the first row of Fig.~\ref{fig:pdffig}.

In Fig.~\ref{fig:pdffig} we present the intrinsic charm PDF from our rigorous expression with three different choices of charm masses. 
The results of the BHPS model, meson cloud model and naive meson cloud model are also juxtaposed for comparison. 
We plot the results of both ground-state and the first excited pions. We find that the profile of our rigorous results significantly differ from the predictions given by the MCM and BHPS models. The results of the naive MCM are $1 \sim 2$ order-of-magnitude smaller than the other results. 
This comparison clearly shows that one can not simply ignore the contribution from the excited charmed meson states when applying the MCM in phenomenological studies. 
It may shed some shadow on the phenomenological work of intrinsic charm PDF in a nucleon~\cite{Melnitchouk:1997ig}.

\begin{figure}[htb]
\includegraphics[width=0.9\textwidth]{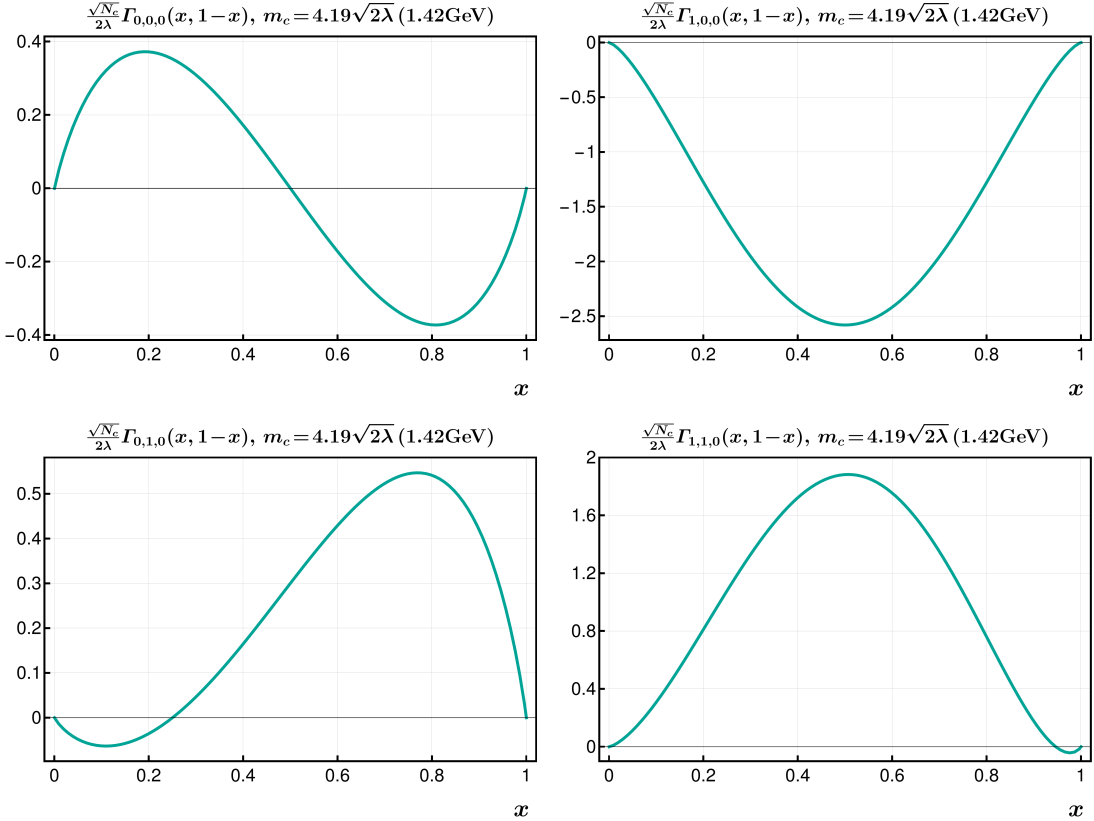}
\caption{The transition vertex function $\varGamma_{n,n_1,n_2}(x,1-x)$. We plot two typical cases, {\it{i.e.}} $n_1=n_2=0$ and $n_1=1$, $n_2=0$ for $n=0$ (left column) and $n=1$ (right column).}
\label{fig:vrtxfunc}
\end{figure}

The intrinsic charm PDF of the first excited $\pi$ is about one order-of-magnitude larger than that of the lowest-lying pion. 
Actually, this can be reflected at the level of transition vertex function. 
In Fig.~\ref{fig:vrtxfunc}, we compare the transition vertex function $\varGamma_{n,n_1,n_2}(x,1-x)$ between the ground state and the first excited state \footnote{Since $\varGamma_{n,n_1,n_2}(x_1,x_2)$ is always accompanied with a $\delta$-function, we only consider the situation that $x_2 = 1- x_1$.}. We observe that the transition vertex function with $n=1$ is significantly larger than $n=0$ case in magnitude.
This difference might be accounted by the distinct charge conjugation properties.
Recall the charge conjugate transformation of the mesonic annihilation operator
\begin{align}
    \mathcal{C}m^{u\bar u}_n(P^+)\mathcal{C}^{-1}
        &=(-1)^{n+1}m^{u\bar u}_n(P^+),
\label{wvcsym}
\end{align}
where $\varphi^{u\bar u}_n(x) = (-1)^{n}\varphi^{u\bar u}_n(1-x)$ has been applied. The ground state and the first excited pion state 
have opposite C-parities. For a pion transitioning into $\ket{D_{n_1}(k^+_1)\overline D_{n_2}(k^+_2)}$, when $n_1=n_2$ and $k^+_1 = k^+_2$, 
the final state has an even C-parity, thus it is only possible if the initial pion is the first-excited state. Correspondingly, $\varGamma_{0,0,0}$ vanishes at $x_1=x_2=\tfrac{1}{2}$. As the vertex function is continuous, the charge conjugate symmetry leads to the suppression of the ground state transition at all $x_1$ as shown in Fig.~\ref{fig:vrtxfunc}.

We also find that when we increase the charm quark mass, the peak position of our result tend to shifting to a larger $x$ value. 
The peak position of MCM remains almost unchanged. In the meanwhile, the magnitude of intrinsic charm PDF from all model predictions 
decrease very fast with increasing charm mass.

To quantitatively investigate how the intrinsic charm PDF depends on the charm quark mass, we also calculate the first two Mellin moments of intrinsic charm PDF:
\begin{equation}
\left\langle x^0 \right\rangle = \int_0^1 dx\, f_c(x),\;\qquad \left\langle x^1 \right\rangle = \int_0^1 dx\, x f_c(x).
\end{equation}
The first two moments have straightforward interpretation: the first moment corresponds to the average number of charm quark inside the pion, while the second moment characterizes 
the average momentum fraction carried by the charm. We vary the charm quark mass ranging from $4.19\sqrt{2\lambda}$ to $40.98\sqrt{2\lambda}$. 
The numerical results of first two Mellin moments are shown in Table~\ref{tab:moments}.

\begin{table}[htb]
\begin{center}
{$n=0$}\\
\vspace{0.5em}
\begin{tabular}{|c|c|c|c|c|c|c|c|c|c|c|}   
 \hline \text{ $m_c [\sqrt{2 \lambda}  (\textcolor{blue}{\text{GeV}})]$}  &  4.19 (\! \textcolor{blue}{1.42} ) &  5.51 (\! \textcolor{blue}{1.87})  &  6.53 (\! \textcolor{blue}{2.22})  &  7.55 (\! \textcolor{blue}{2.57} ) &  8.57 (\! \textcolor{blue}{2.91})  &  9.58 (\! \textcolor{blue}{3.26} )  &  10.60 (\! \textcolor{blue}{3.61} ) &  11.62 (\! \textcolor{blue}{3.95})  \\
\hline  \text{$\left\langle x^0 \right\rangle$} & $ 1.86 \times 10^{-5} $  & $ 6.26  \times 10^{-6} $ & $ 3.09  \times 10^{-6} $  & $ 1.67 \times 10^{-6} $  & $ 9.71  \times 10^{-7} $  & $ 5.78  \times 10^{-7}$  & $ 3.63  \times 10^{-7} $    & $ 2.36  \times 10^{-7} $ \\  
\hline  \text{$\left\langle x^1 \right\rangle$} & $  2.81  \times 10^{-6} $  & $  1.03 \times 10^{-6} $  & $  5.38 \times 10^{-7} $  & $  3.04 \times 10^{-7} $ & $  1.83  \times 10^{-7} $   & $  1.15  \times 10^{-7} $  & $  7.52  \times 10^{-8} $  & $  5.09  \times 10^{-8} $  \\  
\hline  \text{$\left\langle x^1 \right\rangle / \left\langle x^0 \right\rangle$} & 0.151 & 0.164 & 0.174 & 0.183 & 0.189 & 0.200 & 0.207 & 0.216 \\    
\hline  \hline \text{ $m_c [\sqrt{2 \lambda}  (\textcolor{blue}{\text{GeV}})]$}   &  12.64 (\! \textcolor{blue}{4.30})  &  13.66 (\! \textcolor{blue}{4.64} ) &  18.17 (\! \textcolor{blue}{6.18}) &  22.81 (\! \textcolor{blue}{7.76})  &  27.32 (\! \textcolor{blue}{9.29} ) &  32.16 (\! \textcolor{blue}{10.93} ) &  37.00 (\! \textcolor{blue}{12.58} ) &  40.98 (\! \textcolor{blue}{13.93} ) \\
\hline  \text{$\left\langle x^0 \right\rangle$}    & $ 1.58  \times 10^{-7} $  & $ 1.10 \times 10^{-7} $ & $ 2.45  \times 10^{-8} $  & $ 7.12 \times 10^{-9} $  & $ 2.56  \times 10^{-9} $ & $ 9.80 \times 10^{-10} $ & $ 4.24  \times 10^{-10} $  & $ 2.27  \times 10^{-10} $  \\  
\hline  \textbf{$\left\langle x^1 \right\rangle$}    & $  3.50  \times 10^{-8} $   & $  2.51  \times 10^{-8} $ & $ 6.36  \times 10^{-9} $  & $ 2.01 \times 10^{-9} $  & $ 7.74  \times 10^{-10} $  & $ 3.15  \times 10^{-10} $ & $ 1.42  \times 10^{-10} $   & $ 7.86 \times 10^{-11} $   \\  
\hline  \text{$\left\langle x^1 \right\rangle / \left\langle x^0 \right\rangle$}  & 0.222 & 0.230 & 0.259 & 0.283 & 0.303 & 0.320 & 0.335 & 0.346  \\   
\hline   
\end{tabular}   
\\\vskip 1.5em
{$n=1$}\\
\vspace{0.5em}
\begin{tabular}{|c|c|c|c|c|c|c|c|c|}   
 \hline \text{ $m_c [\sqrt{2 \lambda}  (\textcolor{blue}{\text{GeV}})]$}  &  4.19 (\! \textcolor{blue}{1.42} ) &  5.51 (\! \textcolor{blue}{1.87})  &  6.53 (\! \textcolor{blue}{2.22})  &  7.55 (\! \textcolor{blue}{2.57} ) &  8.57 (\! \textcolor{blue}{2.91})  &  9.58 (\! \textcolor{blue}{3.26} )  &  10.60 (\! \textcolor{blue}{3.61} )  &  11.62 (\! \textcolor{blue}{3.95}) \\
\hline  \textbf{$\left\langle x^0 \right\rangle$} & $ 2.37  \times 10^{-4} $  & $ 7.80  \times 10^{-5} $ & $ 3.89 \times 10^{-5} $  & $ 2.14 \times 10^{-5} $  & $ 1.26 \times 10^{-5} $  & $ 7.79 \times 10^{-6}$   & $ 5.05  \times 10^{-6} $  & $ 3.39  \times 10^{-6} $ \\  
\hline  \textbf{$\left\langle x^1 \right\rangle$} & $  5.15  \times 10^{-5} $  & $ 1.74 \times 10^{-5} $  & $  8.81  \times 10^{-6} $  & $  4.92  \times 10^{-6} $ & $  2.95  \times 10^{-6} $   & $  1.86  \times 10^{-6} $    & $  1.22  \times 10^{-6} $ & $  8.35  \times 10^{-7} $  \\ 
\hline  \text{$\left\langle x^1 \right\rangle / \left\langle x^0 \right\rangle$} & 0.218 & 0.223 & 0.226 & 0.230 & 0.235 & 0.238 & 0.242  & 0.247 \\   
\hline  \hline \text{ $m_c [\sqrt{2 \lambda}  (\textcolor{blue}{\text{GeV}})]$}   &  12.64 (\! \textcolor{blue}{4.30})  &  13.66 (\! \textcolor{blue}{4.64} ) &  18.17 (\! \textcolor{blue}{6.18}) &  22.81 (\! \textcolor{blue}{7.76})  &  27.32 (\! \textcolor{blue}{9.29} ) &  32.16 (\! \textcolor{blue}{10.93}) &  37.00 (\! \textcolor{blue}{12.58} ) &  40.98 (\! \textcolor{blue}{13.93} ) \\
\hline  \text{$\left\langle x^0 \right\rangle$}   & $ 2.34  \times 10^{-6} $  & $ 1.64  \times 10^{-6} $ & $ 4.45  \times 10^{-7} $  & $ 1.46  \times 10^{-7} $   & $ 5.76  \times 10^{-8} $ & $ 2.41  \times 10^{-8} $   & $ 1.11  \times 10^{-8} $   & $ 6.23 \times 10^{-9} $    \\  
\hline  \text{$\left\langle x^1 \right\rangle$}  & $  5.87  \times 10^{-7} $   & $  4.20  \times 10^{-7} $ & $ 1.21  \times 10^{-7} $  & $  4.24  \times 10^{-8} $ & $  1.77  \times 10^{-8} $ & $ 7.76 \times 10^{-9} $  & $ 3.72 \times 10^{-9} $ & $ 2.15  \times 10^{-9} $  \\   
\hline  \text{$\left\langle x^1 \right\rangle / \left\langle x^0 \right\rangle$} & 0.251 & 0.256 & 0.273 & 0.291 & 0.307 & 0.322 & 0.335 & 0.345  \\    
\hline   
\end{tabular}   
\caption{The first and second Mellin moments of the intrinsic charm PDF and their ratios. The results of the ground-state and the first excited state pion 
are provided. The charm quark mass are given in unit of $\sqrt{2\lambda}$ and $\mathrm{GeV}$.}\label{tab:moments}
\end{center}
\end{table}

\begin{figure}[htb]
\includegraphics[width=0.9\textwidth]{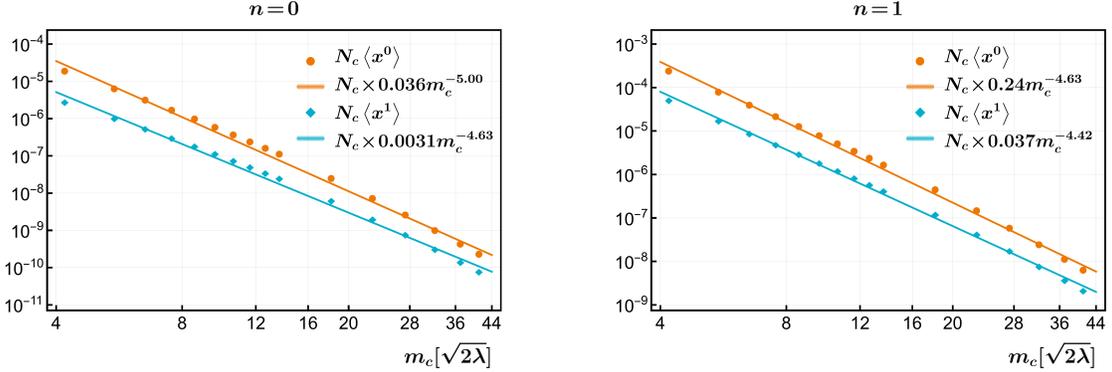}
\caption{The numerical results and the fit of the first two Mellin moments of intrinsic charm PDF with the pion in ground state and the first excited state. }
\label{fig:fitfig}
\end{figure}

We fit the two moments with the simple power-law ansatz $\left\langle x^{0,1} \right\rangle \propto m_c^{d_{0,1}}$. 
The fitting results are shown in Fig.~\ref{fig:fitfig}. We find $d_0=-5.00, d_1=-4.63$ for the ground state and $d_0=-4.63, d_1=-4.42$ for the first excited state. 
The fitted $d_{0,1}$ are pretty close to a naive dimensional analysis prediction $d_{0,1}= -4$ from the meson mass terms in the energy denominator in \eqref{eq:icpdfexp}. 
However, the light-cone wave functions have a rather complicated yet implicit dependence on $m_c$, which may cause the moments to deviate from the $m_c^{-4}$ scaling.

\begin{figure}[htb]
\includegraphics[width=0.6\textwidth]{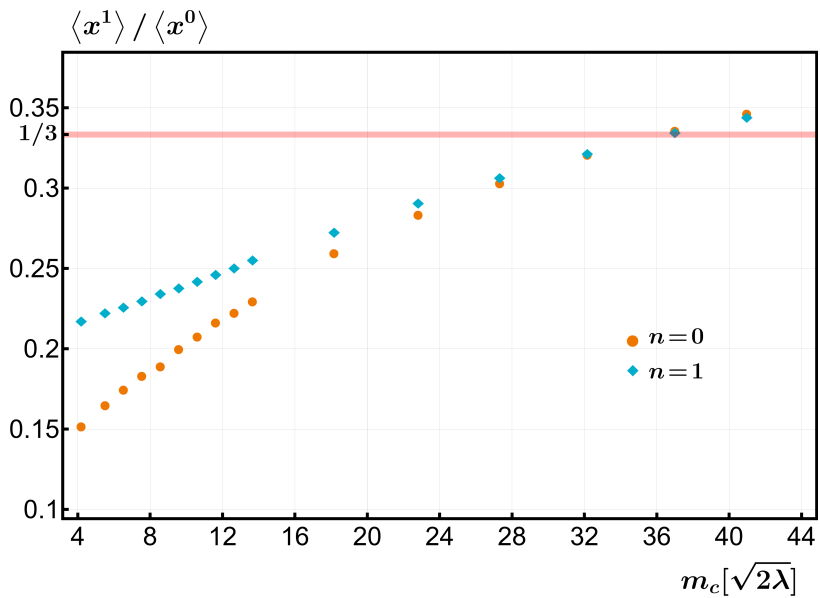}
\caption{The ratio $\langle x^1\rangle/\langle x^0\rangle$ from our rigorous results. The red horizontal line corresponds to the BHPS prediction $\langle x^1\rangle/\langle x^0\rangle=1/3$. Our result surpasses $1/3$ when $m_c\geq 37\sqrt{2\lambda}$. }
\label{fig:ratiofig}
\end{figure}

An interesting finding is that, in the heavy quark limit $m_c \to \infty$, the BHPS model predicts that $\langle x^1\rangle/\langle x^0 \rangle=1/3$ for intrinsic charm~\cite{Brodsky:1981se}, 
while our rigorous results show that the ratio reaches $1/3$ at $m_c\approx 37\sqrt{2\lambda}$ and exceeds the BHPS prediction as the charm quark mass continues to increase. 
We present the numerical results of $\langle x^1\rangle/\langle x^0 \rangle$ in Table~\ref{tab:moments} and Fig.~\ref{fig:ratiofig} for both ground state and the first excited state of pion.

\section{Summary\label{sec_sum}}

Evidence of intrinsic charm PDF of a nucleon has recently aroused tremendous interest in hadron physics community.
In this work, following Collins-Soper's operator definition, we carry out a rigorous study on the intrinsic charm content inside a light neutral meson in the 't Hooft model, {\it i.e.}, the two-dimensional QCD in large-$N_c$ limit.
We explicitly derive the functional form of the intrinsic charm PDF of a light meson
in terms of the 't Hooft wave functions of the light meson and an infinite towers of (anti-)charmed mesons, which first arises at order-$1/N_c$.
For the sake of completeness, we also establish the functional forms of the intrinsic charm PDF predicted by the two-dimensional versions of the BHPS and meson cloud models.
We have made a detailed numerical comparison between our rigorous results and those model predictions.
Especially we notice the close relation between the rigorous result and the MCM prediction, that is, the `interference' terms omitted in the MCM actually have a non-negligible effect on the shape of intrinsic charm PDF.
We also find the contribution from excited charmed hadrons are numerically important, which renders the naive MCM that only considers the lowest-lying charmed hadrons less trustworthy.
Finally, we study how the intrinsic charm PDF of a light meson depends on the charm quark mass.
The numerical studies reveal that the average charm quark number and average momentum fraction carried by the charm quark
in a light meson drop faster than $m_c^{-4}$ as charm quark mass increases.
We hope that our study may shed some light on the nature of the intrinsic charm in the realistic ${\rm QCD}_4$.

\begin{acknowledgments}
The work of S. H., Y. J. and Z. M. is supported in part by the National Natural Science Foundation of China under Grants No. 11925506, 
No. 11621131001 (CRC110 by DFG and NSFC).
The work of X.-N. X. and M.~L. Z. is supported by the National Natural Science Foundation of China under Grants No. 11905296.
\end{acknowledgments}

\end{document}